\documentclass{article}
\usepackage[english]{babel}
\usepackage{amsmath,amssymb,graphicx,tabularx}
\usepackage{xcolor,slashed,epsfig,cite}
\usepackage{verbatim,slashed,ulem,geometry,setspace,algorithm}
\usepackage{algorithmic,mathrsfs,braket,multirow,dcolumn}
\usepackage[colorlinks=true,citecolor=blue,urlcolor=cyan,filecolor=magenta,linkcolor=red,frenchlinks=true,bookmarks]{hyperref} 
\newcommand{\infixand}{\text{ and }}
\newcommand{\nobracket}{}

\newcommand{\tmop}[1]{\ensuremath{\operatorname{#1}}}

\renewcommand\sout{\bgroup \color{red}  \ULdepth=-.5ex \ULset}

\begin{document}

\title{Generalized parton distributions of spin-3/2 particles} 

\author{Dongyan Fu$^{a,b,\thanks{fudongyan@ihep.ac.cn}}$, 
Bao-Dong Sun$^{c,d,e,\thanks{bao-dong.sun@m.scnu.edu.cn}}$,
and Yubing Dong$^{a,b,\thanks{dongyb@ihep.ac.cn}}$\\ 
Institute of High Energy Physics, Chinese Academy 
of Sciences, Beijing 100049, China$^{a}$\\
School of Physical Sciences, University of Chinese 
Academy of Sciences, Beijing 101408, China$^{b}$\\
Guangdong Provincial Key Laboratory of Nuclear Science,
Institute of Quantum Matter,\\
South China Normal University,  Guangzhou 510006, China$^{c}$\\
Guangdong-Hong Kong Joint Laboratory of Quantum Matter,\\
Southern Nuclear Science Computing Center, \\
South China Normal University, Guangzhou 510006, China$^{d}$\\
Helmholtz Institut f\"ur Strahlen- und Kernphysik and Bethe \\
Center for Theoretical Physics, Universit\"at Bonn, D-53115 Bonn, Germany$^{e}$}

\newenvironment{mysubeq}
{\begin{subequations}
\renewcommand\theequation{\theparentequation 
\alph{equation}}}
{\end{subequations}}
\vspace{1em}
\maketitle

\begin{abstract}
Generalized parton distribution functions (GPDs) of spin-3/2 particles 
are defined for the first time in this paper.  Eight unpolarized and eight 
polarized GPDs are found. In the forward limit of GPDs, 
the structure functions and parton distribution functions (PDFs) are obtained. Then, the sum rules that 
connect the GPDs with the electromagnetic and gravitational form 
factors are explicitly displayed. Finally, the relations between GPDs 
and the helicity amplitudes of the system are derived.
\end{abstract}

\section{Introduction}\label{Section1}

\quad The electromagnetic form factors (EMFFs) of Delta 
isobar (spin-3/2) have been explored extensively both experimentally 
and theoretically for a long history~\cite{Weber:1978dh,Benmerrouche:1989uc,
Nozawa:1990gt,Pascalutsa:1998pw,Napsuciale:2006wr,Fu:2022rkn}. 
They show richer information on charge and magnetic structures carried by 
the spin-3/2 particles w.r.t. the lower spin hadrons like 
nucleon~\cite{Gross:2006fg,deMelo:2008rj,Cloet:2014rja,deAraujo:2017uad}, 
deuteron~\cite{Gross:2002ge, Gross:2001ap,Gilman:2001yh,
Garcon:2001sz,Sun:2016ncc,Dong:2008mt} and rho meson~\cite{Krutov:2018mbu,
Krutov:2016uhy,Choi:2004ww,deMelo:1997hh,GarciaGudino:2010sd} etc. 
In parallel, the gravitational form factors (GFFs) which are defined by 
factorizing the matrix element of the energy momentum tensor (EMT) 
characterize the mechanical properties such as the mass, the spin density, 
as well as the internal force 
distributions inside the particles~\cite{Polyakov:2018zvc,Cotogno:2019vjb,
Sun:2020wfo,Kim:2020lrs,Ji:2021mfb,Alharazin:2022wjj}. 
Both EMFFs and GFFs can be related through the GPDs which is firstly 
introduced in describing the deeply virtual Compton scattering (DVCS) 
process for spin-1/2 particles\cite{Muller:1994ses,Radyushkin:1996nd,
Ji:1996nm} and then for spin-1 particles \cite{Berger:2001zb}. GPDs can 
also produce PDFs and structure functions 
in the forward limit. Moreover, through the crossing symmetry, GPDs are 
also connected with generalized distribution amplitudes 
(GDAs)~\cite{Kumano:2017lhr}. It makes the concept of GPDs very 
important to understand the abundant experimental and theoretical (by 
model-(in)dependent studies, lattice QCD calculations etc.) information 
of hadron structures in a unified theoretical framework 
\cite{Diehl:2003ny,Belitsky:2005qn,Diehl:2015uka,Freese:2021mzg}. 
Thus, GPDs and the related quantities have been receiving lots of interest 
for particles of spin-0 (e.g. pion and kaon mesons~\cite{Fanelli:2016aqc,
Zhang:2021tnr,Zhang:2021mtn,Raya:2021zrz,Choi:2001fc,Broniowski:2011xj}), 
spin-1/2 (e.g. nucleon \cite{Sharma:2016cnf,Diehl:2013xca,Selyugin:2015pha,
Gutsche:2016gcd}, and for nuclei, like $^3$He~\cite{Kirchner:2003wt,
Cano:2003ju}) and spin-1 (e.g. rho meson \cite{Sun:2017gtz,Sun:2018ldr,
Kumar:2019eck,Zhang:2022zim} and deuteron \cite{Cano:2003ju,
Kirchner:2003wt,HERMES:2009bqn,Dong:2013rk,Mondal:2017lph}), etc.

In contrast to the increasing interest in the EMFFs and GFFs of 
the Delta isobar (spin-3/2), the definitions of GPDs for spin-3/2 hadrons 
are still missing. One could expect that the spin-3/2 GPDs would expand 
and deepen our understanding of the quantities such as PDFs 
\cite{Chai:2020nxw}, structure functions, EMFFs, and GFFs, etc. 
for higher spin ($\ge 3/2$) particles in a similar method as lower spin 
cases. Therefore it is of great interest to give the definitions and 
properties of spin-3/2 GPDs and it is the main purpose of this work. 
Now the most promising "measurement" is the lattice QCD calculation, 
e.g. the recent lattice calculation on spin-3/2 gluonic 
GFFs\cite{Pefkou:2021fni}. In the future electron-ion collision experiments, 
EIC~\cite{Proceedings:2020eah} (under construction) and 
EicC~\cite{Anderle:2021wcy} (planned), the candidate targets may possess 
higher spins, including spin-3/2 nuclei like $^{7}_{3}$Li, $^{9}_{4}$Be, 
and $^{11}_{4}$B, and even spin-5/2 ones like $^{17}_8$O, $^{25}_{12}$Mg, 
and $^{27}_{13}$Al~\cite{Anderle:2021wcy}. Therefore, it would be possible 
to access the spin-3/2 GPDs and helicity amplitudes experimentally 
\cite{HERMES:2009bqn} in the near future. Nevertheless, 
the spin-3/2 GPDs by themselves are important in the theoretical aspect.

This work is organized as follows:
In Section~\ref{section2}, we give the definitions and properties of GPDs 
of the spin-3/2 system. Section~\ref{section3} displays the connections of 
structure functions and PDFs to GPDs, and two sum rules of the 
structure functions are derived for the spin-3/2 system.
Moreover, the sum rules connecting GPDs with EMFFs and GFFs and the 
helicity amplitudes are expilcitly given. Finally, Section~\ref{sectionsu} 
will be devoted to a summary and discussion.\\

\section{Generalized parton distributions of a spin-3/2 particle}\label{section2}

\begin{figure}[h]
    \centering
    \includegraphics[width=5cm]{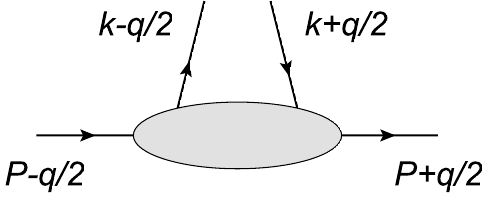} \quad \quad \quad \quad
    \includegraphics[width=5cm]{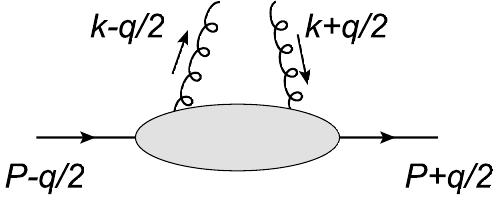}
    \caption{Diagrams describe the GPDs for quarks (left) and gluons (right).}
    \label{fig1}
\end{figure}

\quad The convention of the four-vector 
$v$ in light-cone coordinates is given as
\begin{eqnarray}
v = (v^+, v^-, \mathbf{v}_\perp), \quad 
\tmop{with} \quad v^{\pm} = v^0 \pm v^3
\infixand \mathbf{v}_\perp = (v^1, v^2) , 
\end{eqnarray}
and the light-like four vector $n = (0, 2, \mathbf{0})$ with $n^2=0$. The scalar product of two 
four-vectors is $u \cdot v = \frac{1}{2} u^+ v^- + \frac{1}{2} u^- v^+ - 
\mathbf{u}_\perp \cdot \mathbf{v}_\perp$. The convention for the momenta is
\begin{eqnarray} 
P = \frac{p + p'}{2}, \quad q = p' - p, \quad t = q^2, 
\end{eqnarray}
where $p$ and $p'$ are the initial and final momenta, respectively.
The conventions of variables, skewness $\xi$ and $x$ in the representations of GPDs, are
\begin{equation}
    \xi = - \frac{q \cdot n}{2 P \cdot n} = - 
    \frac{q^+}{2P^+} \quad  (| \xi | \leqslant 1), 
    \quad \text{and} \quad x = \frac{k \cdot n}{P \cdot n} 
    = \frac{k^+}{P^+}
    \quad  (-1 \leqslant x \leqslant 1),
\end{equation}
where $k$ is the loop momentum in Fig.~\ref{fig1}.
It shows that the four-momentum of the parton emitted from the initial particle is $k-q/2$ and the one absorbed in the final particle is $k+q/2$. The corresponding fraction of the momentum carried by the parton over that of the total system in the light cone direction is $x_q=(x+\xi)/(1+\xi) $ for the initial one and $x_q^\prime=(x-\xi)/(1-\xi) $ for the final one \cite{Frederico:2009fk}.
\\

The GPDs are defined through the nondiagonal matrix elements of quark and 
gluon nonlocal current operators at a light-like separation~\cite{Ji:1996ek}.
The general decompositions for quarks can be written as 
\begin{equation}
  \begin{split}
    V_{\lambda' \lambda} & = \frac{1}{2} 
    \int \frac{\text{d} z^-}{2 \pi} e^{i x (P \cdot z)}
    \left.\left\langle p', \lambda'\left| 
    \overline{\psi} \left( - z/2
    \right) \slashed{n} \psi \left(  z/2
    \right)\right| p, \lambda \right\rangle 
    \right|_{z^+ = 0, \mathbf{z} = \mathbf{0}}\\
    & = -\overline{u}_{\alpha'} (p', \lambda')
     \mathcal{H}^{\alpha' \alpha}
    (x, \xi, t) u_{\alpha} (p, \lambda),
 \end{split} \label{unGPDDefine}
\end{equation}
for unpolarized case, and 
\begin{equation}
  \begin{split}
    A_{\lambda' \lambda} & = \frac{1}{2} 
    \int \frac{\text{d} z^-}{2 \pi}e^{i x (P \cdot z)}
    \left. \left\langle p', \lambda' \left| \overline{\psi} 
    \left( -  z/2 \right) \slashed{n} \gamma_5 
    \psi \left(  z/2 \right) \right| p,
    \lambda \right\rangle \right|_{z^+ = 0, \mathbf{z} 
    = \mathbf{0}}\\
    & = - \overline{u}_{\alpha'}(p', \lambda')
    \mathcal{\tilde H}^{\alpha' \alpha}
    (x, \xi, t) u_{\alpha} (p, \lambda),
  \end{split} \label{poGPDDefine}
\end{equation}
for polarized case, where $\lambda \, (\lambda')$ are the helicities of the incoming (outgoing) spin-3/2 particles and $\alpha \, (\alpha')$ are reserved indices for the initial (final) states in this work. The tensors $\mathcal{H}^{\alpha' \alpha}$ and $\mathcal{\tilde H}^{\alpha' \alpha}$ define the GPDs as will be shown in the later context.
The Rarita-Schwinger spinor $u_{\alpha} (p,\lambda)$ shown in~\ref{appendixa} is normalized to $\overline{u}_{\alpha} (p,\lambda') u^{\alpha} 
(p,\lambda) = - 2M \delta_{\lambda' \lambda}$.\\

To count the number of independent GPDs, one can define the helicity amplitudes for the scattering of a quark in a spin-3/2 particle as~\cite{Diehl:2003ny}
\begin{equation} 
  \begin{split}
    \mathcal{A}_{\lambda' \mu', \lambda \mu} 
    =  \int \frac{\text{d} z^-}{2 \pi} 
    e^{i x (P \cdot z)} \left.\langle p', \lambda'\left|
    \mathcal{O}_{\mu' \mu} (z)\right|p,
    \lambda\rangle \right|_{z^+ = 0, \mathbf{z} = \mathbf{0}}
  \end{split} . \label{ha}
\end{equation}
The operators $\mathcal{O}_{\mu' \mu}$ 
describe a quark transferring from helicity $\mu$ to $\mu'$.
The two helicity-conserved operators are 
\begin{equation}\label{operators}
  \begin{split}
    \mathcal{O}_{+ +} (z) & = \frac{1}{4}  \overline{\psi} \left(- z/2 \right)
    \gamma^+  (1 +
    \gamma_5) \psi \left(z/2 \right),\\
    \mathcal{O}_{- -} (z) & = \frac{1}{4}  \overline{\psi} \left(- z/2 \right)
    \gamma^+  (1 -
    \gamma_5) \psi \left(z/2 \right).
  \end{split}
\end{equation}
With those definitions, one has
\begin{equation}\label{VA}
  \mathcal{A}_{\lambda' \pm,\lambda \pm}=\frac{1}{2}(V_{\lambda' \lambda} \pm A_{\lambda' \lambda}),
\end{equation}
where $\pm$ represents the quark helicities. The constraints,
\begin{equation}\label{paritytime}
  \begin{split}
    \mathcal{A}_{- \lambda' - \mu, - \lambda - \mu} 
    = & (- 1)^{\lambda'-\lambda } \mathcal{A}_{\lambda' \mu, \lambda \mu}^\ast,\\
    \mathcal{A}(x,\xi,t)_{ \lambda'  \mu,  \lambda  \mu} 
    = & (-1)^{\lambda' - \lambda} \mathcal{A}^\ast(x,-\xi,t)_{\lambda \mu, \lambda' \mu},
  \end{split}
\end{equation}
can be obtained from the parity and time reversal invariances. And the analogous forms have been found for spin 1/2 and 1 cases in Refs.~\cite{Diehl:2003ny,Berger:2001zb}.
With these constraints, we finally obtain up to eight independent unpolarized GPDs and eight independent polarized GPDs for the spin-3/2 system.\\

The building blocks to construct the Lorentz structures accompanying the spin-3/2 GPDs are: $P^{\mu}$, $q^{\mu}$, $n^{\mu}$, $\gamma^{\mu}$, $\gamma^5$ (or Levi-Civita tensors $\epsilon^{\mu\nu\alpha\beta}$),  
$g^{\mu\nu}$ and $\sigma^{\mu\nu}$.
With help of the on-shell identities given in Refs.~\cite{Cotogno:2019vjb,Fu:2022rkn} and the properties of the Rarita-Schwinger spinor (see~\ref{appendixa}), one can obtain eight independent tensors for the unpolarized and eight for the polarized cases, which agrees with the number-counting from the helicity amplitudes. Equivalently, one can build the tensors accompanying the GPDs of the 
spin-3/2 particle through the direct product between the unpolarized 
spin-1/2 and unpolarized spin-1 structures or the direct product 
between the polarized spin-1/2 and polarized spin-1 structures. 
It can be proven that the ``polarized-polarized'' pairs of tensors 
are equivalent to ``unpolarized-unpolarized'' ones.
It is known that the tensors accompanying the unpolarized GPDs of a spin-1/2 
particle~\cite{Hoodbhoy:1998vm,Diehl:2001pm} have two independent Lorentz structures,
\begin{equation}
    1, \quad \slashed{n},
\end{equation}
and the tensors for the spin-1 case are~\cite{Berger:2001zb,Sun:2017gtz}
\begin{equation}
    g^{\alpha' \alpha}, \quad 
    P^{\alpha'}P^{\alpha}, \quad 
    n^{\left[ \alpha \right.'} P^{\left. \alpha \right]}, \quad 
    n^{\left\{ \alpha \right.'}P^{\left. \alpha \right\}}, \quad 
    n^{\alpha'} n^{\alpha},
\end{equation}
with $a^{[\mu \nobracket} b^{\nobracket \nu]} 
= a^{\mu} b^{\nu} - a^{\nu}b^{\mu}$ and 
$a^{\{ \nobracket \mu} b^{\nu \nobracket \}} = a^{\mu} b^{\nu}
+a^{\nu} b^{\mu}$.
The direct product gives
\begin{equation} \label{unTensor}
  \begin{split}
    &
    g^{\alpha' \alpha}, \quad 
    P^{\alpha'} P^{\alpha}, \quad
    n^{[\nobracket \alpha'} P^{\alpha \nobracket]}, \quad
    n^{\{ \nobracket \alpha'} P^{\alpha \nobracket \}},\quad 
    n^{\alpha'} n^{\alpha},
    \\ &
    g^{\alpha' \alpha}\slashed{n}, \quad
    P^{\alpha'} P^{\alpha} \slashed{n}, \quad
    n^{[\alpha' \nobracket} P^{\nobracket \alpha]}\slashed{n}, \quad
    n^{\{ \nobracket \alpha'} P^{\alpha\nobracket \}}\slashed{n} , \quad 
    n^{\alpha'} n^{\alpha}\slashed{n}.
  \end{split}
\end{equation}
There are two on-shell identities, Eqs.~\eqref{unB1} and \eqref{unB2} (see~\ref{appendixb}), that can be employed to reduce two tensors
$n^{\{ \nobracket \alpha'} P^{\alpha \nobracket \}}$ and $n^{\{ \nobracket \alpha'} 
P^{\alpha \nobracket \}} \slashed{n}$ in terms of the rest eight independent ones. Hence, there are eight independent unpolarized GPDs
for the spin-3/2 case which can be defined as
\begin{equation}\label{unGPDs}
    \begin{split}
     \mathcal{H}^{\alpha' \alpha} & 
     = H_1 \frac{g^{\alpha' \alpha}}{M} + H_2 
     \frac{P^{\alpha'} P^{\alpha}}{M^3} + H_3  
     \frac{n^{[\nobracket \alpha'}
     P^{\alpha \nobracket]}}{M P \cdot n} + H_4  \left[\frac{3 M n^{\alpha'}
     n^{\alpha}}{(P \cdot n)^2} + \frac{g^{\alpha' \alpha}}{M}\right] + H_5  \left[\frac{g^{\alpha'
     \alpha} \slashed{n}}{P \cdot n}-\frac{g^{\alpha' \alpha}}{M}\right]\\
     & + H_6  \frac{P^{\alpha'} P^{\alpha}\slashed{n}}{M^2 P
     \cdot n} + H_7  \frac{n^{[\alpha' \nobracket}
     P^{\nobracket \alpha]} \slashed{n}}{(P \cdot n)^2} 
     + H_8 \left[ \frac{3 M^2 n^{\alpha'} n^{\alpha} 
     \slashed{n}}{(P \cdot n)^3} - \frac{3 M n^{\alpha'}
     n^{\alpha}}{(P \cdot n)^2} \right] .
   \end{split}
\end{equation}\\
where $\mathcal{H}^{\alpha' \alpha} \equiv \mathcal{H}^{\alpha' \alpha}(x,\xi,t)$ and $H_i \equiv H_i(x,\xi,t)$, and similar for the polarized case afterwards.

Analogously, the two independent Lorentz structures of the polarized spin-1/2 GPDs are~\cite{Ji:1996ek}
\begin{equation}
    \gamma^5,\quad \slashed{n}\gamma^5,
\end{equation}
and the four independent tensor structures of the spin-1 polarized GPDs
are~\cite{Berger:2001zb,Sun:2017gtz}
\begin{equation}
    i \epsilon^{n P \alpha' \alpha}, \quad
    i \epsilon^{n P q \{ \alpha \nobracket} P^{\nobracket \alpha' \}} , \quad 
    i \epsilon^{n P q \left[ \alpha \right.} P^{\nobracket \alpha' ] } , \quad 
    i \epsilon^{n P q \{ \alpha \nobracket} n^{\nobracket \alpha' \}} .
\end{equation}
Analogous to the unpolarized case, one expects that the Lorentz structures of the polarized GPDs in the spin-3/2 case can also be expressed 
as the direct product between the polarized and unpolarized structures for the spin-1/2 and spin-1 cases.  The Lorentz structures of 
the polarized GPDs can be written in two equivalent ways: with polarization comes from either spin-1/2 part ($\gamma^5$ terms) or spin-1 part (the Levi-Civita tensors). With the constraints by the on-shell identities (Eqs.~\eqref{poB3} and \eqref{poB4}), there are eight independent polarized GPDs in the spin-3/2 case:
\begin{equation}\label{poGPDs}
    \begin{aligned}
    \tilde{\mathcal{H}}^{\alpha' \alpha}&
    = \tilde{H}_1 \frac{g^{\alpha' \alpha}}{M} \gamma^5 + \tilde{H}_2 
    \frac{P^{\alpha'} P^{\alpha}}{M^3} \gamma^5 
    + \tilde{H}_3  \frac{n^{\left\{ \alpha \right.'}
    P^{\left.\alpha \right\}}}{M P \cdot n}\gamma^5 
    + \tilde{H}_4  \frac{M n^{\alpha'}
    n^{\alpha}}{(P \cdot n)^2} \gamma^5 + \tilde{H}_5  
    \frac{3 g^{\alpha' \alpha}}{\sqrt{5} P \cdot n} \slashed{n} \gamma^5\\
    & + \tilde{H}_6  \frac{3 P^{\alpha'} P^{\alpha}}{\sqrt{5} M^2 \left(P
    \cdot n \right)} \slashed{n} \gamma^5 + \tilde{H}_7  
    \frac{ n^{\left\{ \alpha \right.'}
    P^{\left.\alpha \right\}}}{(P \cdot n)^2}\slashed{n} 
    \gamma^5 + \tilde{H}_8  \left[ \frac{
    \sqrt{5} M^2 n^{\alpha'} n^{\alpha}}{(P \cdot n)^3} +\frac{ g^{\alpha' \alpha}}{\sqrt{5} P \cdot n} \right] \slashed{n} \gamma^5.
    \end{aligned}
\end{equation}\\

The time reversal doesn't provide further limits on the number of GPDs but determines their behavior under the sign change of the skewness parameter $\xi$, 
\begin{subequations}\label{time}
\begin{align}
        H_i (x,\xi,t)=&H_i (x,-\xi,t) 
        \quad \text{with}\quad i=1,2,4,5,6,8,\\
        H_i (x,\xi,t)=& - H_i (x,-\xi,t) 
        \quad \text{with}\quad i=3,7,\\
        \tilde{H}_j (x,\xi,t)=&- \tilde{H}_j (x,-\xi,t) 
        \quad \text{with} \quad j=1,2,3,4,\\
        \tilde{H}_j (x,\xi,t)=& \tilde{H}_j (x,-\xi,t) 
        \quad \text{with} \quad j=5,6,7,8.
\end{align}
\end{subequations}
${H}_{3 , 7}$ and $\tilde{H}_{1, 2,3,4}$ are T-odd GPDs and others are T-even GPDs. When $\xi=0$, ${H}_{3 , 7}(x,0,t) = 0$ and $\tilde{H}_{1, 2,3,4}(x,0,t) = 0$. It should be mentioned that 
The spin-0 and spin-1/2 GPDs are all T-even and the T-odd GPDs start to appear from the spin-1 case which are $H_4^{\text{(S=1)}}$ and ${\tilde{H}}_3^{\text{(S=1)}}$ \cite{Berger:2001zb}.\\

In addition, for the gluon distributions in the spin-3/2 system, 
instead of the matrix elements in \eqref{unGPDDefine} and 
\eqref{poGPDDefine} we have
\begin{subequations}
    \begin{align}
    \frac{n_{\beta'} n_\beta}{P\cdot n} &
    \int \frac{\text{d} z^-}{2 \pi} e^{i x (P \cdot z)}
    \left. \left\langle p', \lambda ' \left|
    F^{\beta' \mu}\left(-z/2\right)
    F_{\mu}^{~\beta}\left(z/2\right)
    \right| p, \lambda \right\rangle \right|_{z^+ = 0, 
    \mathbf{z} = \mathbf{0}} \notag \\
    & = - \overline{u}_{\alpha'} (p', \lambda') \mathcal{H}^{\alpha'
    \alpha}_g (x, \xi, t) u_{\alpha} (p, \lambda), \\
    -i \frac{n_{\beta'} n_\beta}{P\cdot n} &
    \int \frac{\text{d} z^-}{2 \pi} e^{i x (P \cdot z)}
    \left. \left\langle p', \lambda ' \left|
    F^{\beta' \mu}\left(- z/2\right)\, 
    \tilde{F}_{\mu}{}^\beta\left( z/2\right)
    \right| p, \lambda \right\rangle \right|_{z^+ = 0, 
    \mathbf{z} = \mathbf{0}} \notag \\
    & = - \overline{u}_{\alpha'} (p', \lambda')
    \tilde{\mathcal{H}}^{\alpha' \alpha}_g (x, \xi, t) 
    u_{\alpha} (p, \lambda),
    \end{align}
\end{subequations}
with $\tilde{F}^{\alpha\beta}=\frac{1}{2}
\epsilon^{\alpha\beta\gamma\delta} F_{\gamma\delta}$. The tensors 
$\mathcal{H}^{\alpha' \alpha}_g$, $\tilde{\mathcal{H}}^{\alpha' \alpha}_g$ 
have the same structures as those for quark distributions given in 
Eqs.~\eqref{unGPDs} and~\eqref{poGPDs}.
It should be stressed that Diehl's convention \cite{Diehl:2003ny} is used 
here, and the definitions of gluon GPDs under Ji's convention
\cite{Ji:1998pc} would differ by a factor $2 x$, i.e.,  $H_g = 2 x 
H_{g}^{(\text{Ji})}$ and $\tilde{H}_g = 2 x \tilde{H}_{g}^{(\text{Ji})}$.

\section{PDFs, sum rules and helicity amplitudes}\label{section3}

\subsection{The forward limit}

\quad It is known that the GPDs in the forward limit give the usual parton 
distribution functions. In the parton model for the spin-3/2 sector, 
there are four independent structure functions in deep inelastic scattering 
at leading twist and leading order in $\alpha_s$. They are 
$F_1$, $b_1$, $g_1$, $g_2$ whose probabilistic interpretations in terms 
of quark densities read~\cite{Jaffe:1988up}
\begin{subequations}\label{structurefunctions}
  \begin{align}
    F_1(x) =& \frac{1}{2} \sum_q e_q^2 
    \frac{q^{\frac{3}{2}}_{\uparrow} (x) + q^{-
    \frac{3}{2}}_{\uparrow} (x) + q^{\frac{1}{2}}_{\uparrow} (x) 
    + q^{-\frac{1}{2}}_{\uparrow} (x)}{2} 
    + \{ q \rightarrow \overline{q} \},\\
    b_1(x) =& \frac{1}{2} \sum_q e_q^2 
    \frac{\left( q^{\frac{3}{2}}_{\uparrow} (x)
    + q^{- \frac{3}{2}}_{\uparrow} (x) \right) - \left(
    q^{\frac{1}{2}}_{\uparrow} (x) + q^{- \frac{1}{2}}_{\uparrow} (x)
    \right)}{2} + \{ q \rightarrow \overline{q} \},\\
    g_1(x) =& \frac{1}{2} \sum_q e_q^2 
    \frac{3 \left( q^{\frac{3}{2}}_{\uparrow}
    (x) - q^{- \frac{3}{2}}_{\uparrow} (x) \right) + \left(
    q^{\frac{1}{2}}_{\uparrow} (x) - q^{- \frac{1}{2}}_{\uparrow} (x)
    \right)}{\sqrt{20}} + \{ q \rightarrow \overline{q} \},\\
    g_2(x) =& \frac{1}{2} \sum_q e_q^2 
    \frac{\left( q^{\frac{3}{2}}_{\uparrow} (x)
    - q^{- \frac{3}{2}}_{\uparrow} (x) \right) - 3 \left(
    q^{\frac{1}{2}}_{\uparrow} (x) - q^{- \frac{1}{2}}_{\uparrow} (x)
    \right)}{\sqrt{20}} + \{ q \rightarrow \overline{q} \} ,
  \end{align}
\end{subequations}
where $q^\lambda_{\uparrow}(x)$ stands for the probability to find a quark with momentum fraction $x$ and positive helicity in the spin-3/2 particle with helicity $\lambda$. In addition, one 
has $q^\lambda_{\uparrow}(x) = q^{- \lambda}_{\downarrow}(x)$ 
from parity invariance. Here, $g_2$ is the new structure function as the spin goes from 1 up to 3/2. In the forward limit, there are $\Bar{u}_{\alpha'} P^{\alpha'}=P^\alpha u_\alpha=0$, so the only structures in Eqs. (\ref{unGPDs}) and (\ref{poGPDs}) that survive are those proportional 
to $H_i$ and $\tilde{H}_i$ with $i=1,4,5,8$.
Moreover, in the forward limit, $\tilde{H}_{1,4}$ vanish because of the time reversal relation \eqref{time} and $H_{5,8}$ vanish as well because of $\Bar{u}_{\alpha'} (M \slashed{n}-P \cdot n) u_\alpha=0$.
According to the results for the helicity amplitudes shown below, one gets
\begin{subequations}\label{sf0}
  \begin{align}
    2 H_1 (x,0,0)=&
    \frac{q^{\frac{3}{2}}_{\uparrow} (x) 
    + q^{- \frac{3}{2}}_{\uparrow} (x) +
    q^{\frac{1}{2}}_{\uparrow} (x) 
    + q^{- \frac{1}{2}}_{\uparrow} (x)}{2},\\
    2 H_4 (x,0,0)=& \frac{\left(
    q^{\frac{3}{2}}_{\uparrow} (x) + 
    q^{- \frac{3}{2}}_{\uparrow} (x) \right)
    - \left( q^{\frac{1}{2}}_{\uparrow} (x) 
    + q^{- \frac{1}{2}}_{\uparrow} (x)\right)}{2},\\
    2 \tilde{H}_5 (x,0,0) =& \frac{3
    \left( q^{\frac{3}{2}}_{\uparrow} (x) 
    - q^{- \frac{3}{2}}_{\uparrow} (x)
    \right) + \left( q^{\frac{1}{2}}_{\uparrow} (x) - q^{-
    \frac{1}{2}}_{\uparrow} (x) \right)}{\sqrt{20}},\\
    2 \tilde{H}_8(x,0,0)
    =& \frac{\left( q^{\frac{3}{2}}_{\uparrow} (x) 
    -q^{- \frac{3}{2}}_{\uparrow} (x) \right) 
    - 3 \left(q^{\frac{1}{2}}_{\uparrow} (x) 
    - q^{- \frac{1}{2}}_{\uparrow} (x) \right)}{\sqrt{20}},
  \end{align}
\end{subequations}
for $x>0$. Similar to the deuteron case \cite{Berger:2001zb}, the corresponding relations for $x < 0$ involve the antiquark distributions at $-x$, with an overall minus sign in the expressions for $H_1$ and $H_4$.
With the sum rules \eqref{sumrules} given in the next subsection for the GPDs, the structure functions have the following sum rules,
\begin{equation}\label{sf}
    \int ^1_0 \text{d} x \, b_1(x)= 0, \quad
    \int ^1_0 \text{d} x \, g_2(x)= 0,
\end{equation}
if the quark sea $q-\Bar{q}$ does not contribute to the integral. These two equalities in Eq.~\eqref{sf} are consistent with the sum rules derived from the rotation properties of the structure functions \cite{Jaffe:1988up}.

\subsection{Sum rules}

\quad As shown in Ref.~\cite{Diehl:2003ny}, the $(a+1)$th Mellin moments (in $x$) of the operator defining the quark GPDs of the 
system~\eqref{unGPDDefine} lead to derivative operators between the two fields, 
\begin{equation}
  \begin{aligned}
    & (P \cdot n)^{a + 1} \int \text{d} x \, x^a \int \frac{\text{d} z^-}{2 \pi} 
    e^{i x P^+ z^-} \left[\, \overline{\psi}
    \left( - z/2 \right) \slashed{n}
    \psi \left( z/2 \right) \right]
    \Bigg |_{z^+ = 0, \mathbf{z} 
    = 0}\\
    = & \left( i \frac{\text{d}}{\text{d} z^-} \right)^a \left[\, \overline{\psi}
    \left( - z/2 \right) \slashed{n} 
    \psi \left( z/2
    \right) \right] \Bigg |_{z = 0} = \overline{\psi} (0) 
    \slashed{n} (i\overleftrightarrow{\partial}^+)^a \psi (0) .
  \end{aligned} \label{Mellin}
\end{equation}
This relation at the operator level connects the quark GPDs with EMFFs 
($a = 0$), GFFs ($a = 1$), and other FFs from higher rank current operators. For gluon GPDs, there exist similar relations as shown explicitly by Ref.~ \cite{Diehl:2003ny}.\\

The decompositions of the matrix elements of the vector~\cite{Nozawa:1990gt} and axial vector~\cite{Alexandrou:2010tj,Jun:2020lfx} currents of the spin-3/2 case are: 
\begin{equation}
    \begin{aligned}
      \langle p',\lambda ' \left| \Bar{\psi}(0) \gamma^\mu 
      \psi (0) \right| p, \lambda \rangle
      = & -2\overline{u}_{\alpha'} (p', \lambda') 
      \left[ g^{\alpha' \alpha} \left( G_1 (t) \frac{P^{\mu}}{M} 
      + G_5 (t) \gamma^{\mu} \right) \right. \\
     & \left. + \frac{P^{\alpha'} P^{\alpha}}{M^2} 
     \left( G_2 (t) \frac{P^{\mu}}{M} + G_6 (t) 
     \gamma^{\mu} \right) \right] u_{\alpha} (p, \lambda),
    \end{aligned}\label{vectord}
\end{equation}
\begin{equation}
    \begin{aligned}
      \langle p',\lambda ' \left| \Bar{\psi}(0) 
      \gamma^\mu \gamma^5 \psi (0) \right| p, \lambda \rangle
     = & -2\overline{u}_{\alpha'} (p', \lambda') 
     \left[ g^{\alpha' \alpha} \left(- \tilde{G}_1 (t) \frac{q^{\mu}}{2M} 
      + \tilde{G}_5 (t) \gamma^{\mu} \right) \right. \\
     & \left. + \frac{P^{\alpha'} P^{\alpha}}{M^2} 
     \left(- \tilde{G}_2 (t) \frac{q^{\mu}}{2M} + \tilde{G}_6 (t) 
     \gamma^{\mu} \right) \right] \gamma^5 u_{\alpha} (p, \lambda),
    \end{aligned}\label{axiald}
\end{equation}
where a different set of notations is adopted for later convenience to exhibit  their relations with the quark GPDs. And the relations between the different notations are $ 2 \left( G_1, G_2, G_5, G_6 \right)=\left( -a_2, c_2, -a_1, c_1 \right)$~\cite{Nozawa:1990gt} and $ 2 \left( \tilde{G}_1, \tilde{G}_2, \tilde{G}_5, \tilde{G}_6 \right) = \left( -g_3, h_3, g_1, -h_1 \right)$~\cite{Jun:2020lfx}. 
Note that the matrix elements (as well as tensors $\mathcal{H}^{\alpha' \alpha}$, $\mathcal{\tilde H}^{\alpha' \alpha}$ and GPDs) are defined flavor by flavor, so one should multiply the electric or weak charges and sum over flavors to get the conventional form factors. The isospin symmetry is not specified in this work.

The GFFs for the spin-3/2 particle are defined as~\cite{Cotogno:2019vjb,Kim:2020lrs}
\begin{equation}
  \begin{aligned}
    & \left\langle p', \lambda' \left| \hat{T}^{\mu \nu} (0) \right| p, \lambda \right\rangle\\
    = & - \overline{u}_{\alpha'} (p', \lambda')
    \left[ \frac{P^{\mu} P^{\nu} }{M}
    \left( g^{\alpha' \alpha} F^T_{1, 0} (t) + \frac{2 P^{\alpha'}
    P^{\alpha}}{M^2} F_{1, 1}^T (t) \right) \right.\\
    & + \frac{(q^{\mu} q^{\nu} - g^{\mu \nu} q^2)}{4 M}  
    \left( g^{\alpha'
    \alpha} F^T_{2, 0} (t) + \frac{2 P^{\alpha'} 
    P^{\alpha}}{M^2} F_{2, 1}^T(t) \right)\\
    & + M g^{\mu \nu} \left( g^{\alpha' \alpha} F^T_{3, 0} (t) 
    + \frac{2 P^{\alpha'} P^{\alpha}}{M^2} F_{3, 1}^T (t) 
    \right) + \frac{P^{\{ \mu
    \nobracket} i \sigma^{\nobracket \nu \} q}}{2 M}  
    \left( g^{\alpha' \alpha}
    F^T_{4, 0} (t) + \frac{2 P^{\alpha'} P^{\alpha}}{M^2} 
    F_{4, 1}^T (t)\right)\\
    & \left. - \frac{1}{M} \left(2 q^{ \{ \mu \nobracket} 
    g^{\nobracket \nu \} [ \alpha' \nobracket}
    P^{\nobracket \alpha ]} + 8 g^{\mu \nu} P^{\alpha'} 
    P^{\alpha} - g^{\alpha'
    \{ \mu \nobracket} g^{\nobracket \nu \} \alpha} q^2 \right) 
    F^T_{5, 0} (t) + M g^{\alpha' \{ \mu \nobracket} 
    g^{\nobracket \nu \} \alpha} 
    F^T_{6, 0}(t) \right]u_{\alpha} (p, \lambda).
  \end{aligned} \label{GFFs}
\end{equation}

The individual tensors in Eq.~\eqref{unTensor} make it more convenient for establishing the polynomiality sum rules. It's similar for the polarized case. Therefore, we introduce another set of coefficient functions that accompany the individual tensors,
\begin{equation} \label{coeff_E}
\begin{split}
    E_1 = & H_1 + H_4 - H_5, \quad E_4 = 3 H_4 - 3 H_8, \quad E_8 = 3 H_8,\\
    E_i = & H_i \quad \text{with} \quad i = 2,3,5,6,7,
\end{split}
\end{equation}
and
\begin{equation}\label{coeff_Et}
\begin{split}
    \tilde{E}_5=&\frac{3}{\sqrt{5}} \tilde{H}_5 + \frac{1}{\sqrt{5}}\tilde{H}_8, \quad \tilde{E}_6=\frac{3}{\sqrt{5}} \tilde{H}_6, \quad \tilde{E}_8=\sqrt{5} \tilde{H}_8,\\
     \quad \tilde{E}_j =& \tilde{H}_j \quad \text{with} \quad j = 1,2,3,4,7.
\end{split}
\end{equation}
Clearly, $E_i$'s ($\tilde{E}_i$'s) are linear combinations of $H_i$'s ($\tilde{H}_i$'s) that possessing the same symmetry in Eq. \eqref{time}.\\

Taking $a=0$ in Eq.~\eqref{Mellin} it gives the sum rules connecting quark GPDs with EMFFs,
\begin{subequations}\label{sumrules}
    \begin{align}
     \int ^1_{-1}\text{d} x \, E_i (x,\xi,t)=& G_i(t) \quad \text{with} 
      \quad i=1,2,5,6,\\
     \int ^1_{-1} \text{d} x \, \tilde{E}_i (x,\xi,t)=& \xi \tilde{G}_i(t) 
      \quad \text{with} \quad i=1,2,\\
     \int ^1_{-1} \text{d} x \, \tilde{E}_i (x,\xi,t)=& \tilde{G}_i(t) 
      \quad \text{with} \quad i=5,6,\\
     \int ^1_{-1}\text{d} x \, E_j (x,\xi,t)= & \int ^1_{-1}\text{d} x \, \tilde{E}_j (x,\xi,t) = 0 \quad \text{with} \quad j=3,4,7,8.
    \end{align}
\end{subequations}
The first moments of $E_3$ and $\tilde{E}_3$ vanish because of the time reversal as shown in Eq.~\eqref{time}. Moreover, the first moments of $E_i\; (i=4,7,8)$ and $\tilde{E}_i\; (i=4,7,8)$ also disappear due to that the tensor structures $n^\mu n^\nu/(P \cdot n)^2$ and $n^\mu n^\nu n^\rho/(P \cdot n)^3$ have no correspondences in the factorization of the matrix elements of local currents.\\

A similar procedure can be done for the case of GFFs. Taking $a=1$ in Eq.~\eqref{Mellin} it gives the sum rules connecting quark GPDs with GFFs,
\begin{subequations}
  \begin{align}
    \int ^1_{-1}\text{d} x \, x E_1 (x, \xi, t) =&  F_{1, 0}^T (t) 
    +  \xi^2 F_{2, 0}^T (t) - 2F_{4, 0}^T (t),\\
    \int ^1_{-1}\text{d} x \, x E_2 (x, \xi, t) =& 2 F_{1, 1}^T (t) 
    + 2 \xi^2 F_{2, 1}^T(t) - 4F_{4, 1}^T (t),\\
    \int ^1_{-1}\text{d} x \,x E_3 (x, \xi, t) =& 8 \xi F_{5, 0}^T (t),\\
    \int ^1_{-1}\text{d} x \,x E_4 (x, \xi, t) =& \frac{2 t}{M^2} F^T_{5, 0} (t) + 2 F^T_{6, 0} (t),\\
    \int ^1_{-1}\text{d} x \,x E_5 (x, \xi, t) =& 2 F^T_{4, 0} (t),\\ 
    \int ^1_{-1}\text{d} x \,x E_6 (x, \xi, t) =& 4 F_{4, 1}^T (t),\\
    \int ^1_{-1}\text{d} x \,x E_i (x, \xi, t) =& 0 \quad \text{with} \quad i = 7, 8.
  \end{align} \label{GFFs0}
\end{subequations}
The second Mellin moment of $E_7$ vanishes because of the time reversal invariance and that of $E_8$ disappears as well because the tensor 
structure $n^\mu n^\nu n^\rho/(P \cdot n)^3$ does not have the 
correspondence in the parametrization of GFFs in Eq.~\eqref{GFFs}.
For the gluon GPDs, the factor $x$ should be absent and the 
integral should only go from 0 to 1.

\subsection{Helicity amplitudes}\label{GPDothers}

\quad The helicity amplitudes given in Eq.~(\ref{ha}) can be expressed in terms of the obtained GPDs. To show the symmetry properties carried by the individual tensor structures, we again express helicity amplitudes in terms of the coefficient functions $E$'s in Eq.~\eqref{coeff_E} and ${\tilde E}$'s in \eqref{coeff_Et} instead of GPDs directly. We introduce the notations
$|\mathbf{p}_{\perp}|e^{\pm i\phi} \equiv p^1\pm i p^2 $ and $|\mathbf{p}'_{\perp}|e^{\pm i\phi'} \equiv p'\,^1\pm i p'\,^2$, and 
\begin{equation}
  \begin{split}
    & C \equiv \sqrt{\frac{1-\xi}{1+\xi}} \frac{| \mathbf{p}_{\perp} |}{M} e^{- i \phi} - \sqrt{\frac{1+\xi}{1-\xi}} 
    \frac{| \mathbf{p}'_{\perp} |}{M} e^{- i \phi'},\\
    & D\equiv -\frac{t}{4 M^2} -\frac{\xi^2}{1-\xi^2},\\
    & K_{\pm i} \equiv   A_{1} E_i \pm \xi A_{1} \tilde{E}_i + A_{2} \tilde{K}_{\pm (i+4)} \quad \text{with} \quad i=1\sim 4,\\
    & \tilde{K}_{\pm j} \equiv   E_j \pm \tilde{E}_j \quad \text{with} \quad j=1\sim 8,
  \end{split}
\end{equation}
where
\begin{equation}
  \begin{split}
    A_{1} \equiv  \frac{2}{\sqrt{1 - \xi^2}}, \quad A_{2} \equiv  2 \sqrt{1 - \xi^2}.
  \end{split}
\end{equation}
Then the helicity amplitudes have the following forms,
\begin{equation}\label{ap3p3}
  \begin{split}
    2\mathcal{A}_{\frac{3}{2} +, \frac{3}{2} +}  = & K_{+ 1} + \frac{| C |^2}{8} K_{+ 2},
  \end{split}
\end{equation}

\begin{equation}\label{ap3p1}
  \begin{split}
    2\mathcal{A}_{\frac{3}{2} +, \frac{1}{2} +} = & -
    \sqrt{\frac{1 + \xi}{1 - \xi}} \frac{C}{\sqrt{3}} \left(
    K_{+ 1} - \frac{1 + \xi}{2} K_{- 3} \right) -
    \frac{C}{\sqrt{3}} \left( \tilde{K}_{- 1} +
    \frac{| C |^2}{8} \tilde{K}_{- 2} \right) \\
    & - \sqrt{\frac{1 + \xi}{1 - \xi}} \frac{ \left[D (1 - \xi^2) + \xi\right] C}{2 \sqrt{3} (1 - \xi^2)} K_{+ 2},
  \end{split}
\end{equation}

\begin{equation}
  \begin{split}
    2\mathcal{A}_{\frac{3}{2} +, (- \frac{1}{2}) +} = & \sqrt{\frac{1 + \xi}{1
    - \xi}} \frac{C^2}{ \sqrt{3}} \left( \tilde{K}_{- 1} - \frac{1 +
    \xi}{2} \tilde{K}_{+ 3} \right) - \frac{C^2}{8 \sqrt{3}} K_{+ 2}
    + \sqrt{\frac{1 + \xi}{1 - \xi}} \frac{\left[D (1 - \xi^2) + \xi\right]
    C^2}{2 \sqrt{3} (1 - \xi^2)} \tilde{K}_{- 2},
  \end{split}
\end{equation}

\begin{equation}
  2\mathcal{A}_{\frac{3}{2} +, (- \frac{3}{2}) +} = \frac{C^3}{8} \tilde{K}_{- 2},
\end{equation}

\begin{equation}\label{ap1p3}
  \begin{split}
    2\mathcal{A}_{\frac{1}{2} +, \frac{3}{2} +} = & \sqrt{\frac{1 - \xi}{1
    + \xi}} \frac{C^{\ast}}{\sqrt{3}} \left( K_{+ 1} + \frac{1 - \xi}{2}
    K_{+ 3} \right) + \frac{C^{\ast}}{\sqrt{3}} \left( \tilde{K}_{+ 1} +
    \frac{| C |^2}{8 M^2} \tilde{K}_{+ 2} \right) \\
    & + \sqrt{\frac{1 - \xi}{1 + \xi}} \frac{[D (1 - \xi^2) - \xi]
    C^{\ast}}{2 \sqrt{3} (1 - \xi^2)} K_{+ 2},
  \end{split}
\end{equation}

\begin{equation}\label{ap1p1}
  \begin{split}
    2\mathcal{A}_{\frac{1}{2} +, \frac{1}{2} +} = & - \frac{2}{3}  \left[
    K_{+ 1} - \frac{1}{2} \left( K_{- 1} + \frac{| C |^2}{8} K_{- 2}
    \right) + K_{+ 3} - K_{- 3} + (1 - \xi^2) K_{+ 4} \right]\\
    & - \frac{| C |^2}{3} \left[ \sqrt{\frac{1 + \xi}{1 - \xi}}
    \tilde{K}_{+ 1} + \sqrt{\frac{1 - \xi}{1 + \xi}} \tilde{K}_{- 1} \right] +
    \frac{| C |^2}{6 \sqrt{1 - \xi^2}} (\tilde{K}_{+ 2} + \tilde{K}_{- 2}
    + 4 \xi \tilde{K}_{- 3})\\
    & + \frac{2 [D (\xi^2 - 1) + 1]}{3} \left( \frac{2 K_{+ 1}}{(1 -
    \xi^2)} + \frac{K_{+ 3}}{1 + \xi} - \frac{K_{- 3}}{1 - \xi} \right)
    - \frac{2 [D^2 (1 - \xi^2)^2 - \xi^2]}{3 (1 - \xi^2)^2} K_{+ 2}\\
    & + \frac{ | C |^2 [D (\xi^2 - 1) - 1]}{6 (1 - \xi^2)} \left(
    \sqrt{\frac{1 + \xi}{1 - \xi}} \tilde{K}_{+ 2} + \sqrt{\frac{1 - \xi}{1 +
    \xi}} \tilde{K}_{- 2} \right),
  \end{split}
\end{equation}

\begin{equation}
  \begin{split}
    2\mathcal{A}_{\frac{1}{2} +, (- \frac{1}{2}) +} = & \frac{2 C}{3}
    \left[ \tilde{K}_{- 1} - \frac{| C |^2}{16} \tilde{K}_{+ 2} -
    \tilde{K}_{+ 3} + \tilde{K}_{- 3}+(1 - \xi^2) \tilde{K}_{- 4} \right] \\
    & - \frac{C}{3} \left( \sqrt{\frac{1 + \xi}{1 - \xi}} K_{- 1} + \sqrt{\frac{1 - \xi}{1 + \xi}} K_{+ 1} \right)
    + \frac{C}{6 \sqrt{1 - \xi^2}} (K_{+ 2} + K_{- 2} + 4 \xi K_{+ 3}) \\
    & - \frac{2 [D (\xi^2 - 1) + 1] C}{3} \left( \frac{2 \tilde{K}_{-
    1}}{(1 - \xi^2)} - \frac{\tilde{K}_{+ 3}}{1 - \xi} + \frac{\tilde{K}_{-
    3}}{1 + \xi} \right) + \frac{2 [D^2 (\xi^2 - 1)^2 - \xi^2] C}{3 (1 - \xi^2)^2} \tilde{K}_{-2}\\
    & + \frac{ [D (\xi^2 - 1) - 1] C}{6 (1 - \xi^2)} \left(
    \sqrt{\frac{1 + \xi}{1 - \xi}} K_{- 2} + \sqrt{\frac{1 - \xi}{1 + \xi}}
    K_{+ 2} \right),
  \end{split}
\end{equation}

\begin{equation}\label{ap1m3}
  \begin{split}
    2\mathcal{A}_{\frac{1}{2} +, (- \frac{3}{2}) +} = & \sqrt{\frac{1 -
    \xi}{1 + \xi}} \frac{C^2}{\sqrt{3}} \left( \tilde{K}_{- 1} + \frac{1
    - \xi}{2} \tilde{K}_{- 3} \right) - \frac{C^2}{8 \sqrt{3}} K_{- 2}
    - \sqrt{\frac{1 - \xi}{1 + \xi}} \frac{[D (\xi^2 - 1) + \xi] C^2}{2 \sqrt{3} (1 - \xi^2)} \tilde{K}_{- 2}.
  \end{split}
\end{equation}

The rest helicity amplitudes can be obtained through the relations in Eqs.~\eqref{VA} and \eqref{paritytime}.
Obviously, $\mathcal{A}_{\frac{3}{2} +, \frac{3}{2} +}$ and $\mathcal{A}_{\frac{1}{2} +, \frac{1}{2} +}$ in Eqs.~\eqref{ap3p3} and \eqref{ap1p1} are unchanged after the time reversal. 
Noted that $C$ is a dimensionless complex number and depends on the transverse momenta of both initial and final states. Thus, the time reversal could turn $C$ into its complex conjugate. It should be addressed that a  common factor $C^{\lambda'-\lambda} \theta(\lambda'-\lambda)+C^{\ast\lambda -\lambda'}\theta(\lambda-\lambda')$, where $\theta(x-y)$ is the step function, can be extracted out of the helicity amplitude $A_{\lambda' +, \lambda +}$ when $\lambda' \neq \lambda$.
In the forward limit, the factor $C$ vanishes and so for $A_{\lambda' +, \lambda +}$'s when $\lambda'\neq \lambda$.
Furtherly, we have $\mathcal{A}(x,0,0)_{\lambda +, \lambda +}=q^\lambda_{\uparrow}(x)$ and $\mathcal{A}(x,0,0)_{\lambda -, \lambda -}=q^\lambda_{\downarrow}(x)$, using which we find the relations in Eq.~\eqref{sf0}.

\section{Summary and discussions}\label{sectionsu}

\quad
In this work, the GPDs of the spin-3/2 particle are given for the first time. There are eight independent unpolarized GPDs and eight polarized ones.
The independent tensors that accompany the distributions can be constructed through the direct product between the tensors accompanying the spin-1/2 unpolarized or polarized GPDs and those of the spin-1.
Moreover, the structure functions can be expressed as the GPDs in the forward limit, and the sum rules connecting GPDs with EMFFs and GFFs are obtained through the Mellin moments.
In the last subsection, the helicity amplitudes of the spin-3/2 particle are derived and expressed in terms of the coefficient functions
which are linear combinations of the GPDs and they depend on the 
transverse momenta of the initial and final states. The parity 
and time reversal invariances are satisfied throughout.
It is expected that the relations given in this work could be tested in the electron-ion collision experiments at future EIC and EicC with spin-3/2 
targets, such as $^7_3$Li. Other measurements about the GPDs of the 
spin-3/2  particle, like the $\Omega$ baryon, may also be possible in 
the  heavy ion collision, where the $\Omega$ baryon is  rich in the final state. Finally, the numerical results for the unpolarized and polarized 
GPDs of a spin-3/2 particle, taking the $\Delta$ isobar as an example, will be given in our forthcoming work.

\section*{Acknowledgments}
\quad The authors are grateful to Wim Cosyn for the helpful discussion on the forms of spinor and polarization vector. 
This work is supported by the National Natural 
Science Foundation of China under Grants No. 11975245, 
No. 11947224, No. 11947228, and No. 12035007. This work is 
also supported by the Sino-German CRC 110 “Symmetries and the 
Emergence of Structure in QCD” project by NSFC under Grant
No. 12070131001, the Key Research Program of Frontier Sciences, 
CAS, under Grant No. Y7292610K1, and the National Key Research 
and Development Program of China under Contracts 
No. 2020YFA0406300, and Guangdong Provincial funding with 
Grant No. 2019QN01X172, 
Guangdong Major Project of
Basic and Applied Basic Research No. 2020B0301030008, and the Department of Science and Technology of Guangdong Province with Grant No. 2022A0505030010.

\appendix
\renewcommand\thesection{Appendix~\Alph{section}}
\section{The Rarita-Schwinger spinor}\label{appendixa}
\vspace{0.2cm} \par\noindent\par\setcounter{equation}{0}
\renewcommand{\theequation}{A\arabic{equation}}

The explicit form of the Rarita-Schwinger spinor of a spin-3/2 particle employed in
our work is \cite{Rarita:1941mf}
\begin{equation}\label{RSspinor}
u^{\alpha} (p, \lambda) = \sum_{\rho,\sigma} C^{\frac{3}{2} \lambda}_{1 \rho, 
\frac{1}{2}\sigma} \epsilon^{\alpha} (p, \rho) u (p, \sigma),
\end{equation}
where the coefficient in \eqref{RSspinor} is the 
Clebsch-Gordan coefficient. The explicit light-front form expressions of the polarization vectors are derived from Ref.~\cite{Keister:1991sb} by boosting from the rest momentum frame to the moving frame,
\begin{subequations}\label{vectora2}
    \begin{align}
      \epsilon^{\alpha} (p,0) & = \frac{1}{M} \left(
      p^+,p^- - \frac{2 M^2}{p^+},\boldsymbol{\epsilon}_\perp (p,0) \right)^\text{T} \quad \text{with} \quad \boldsymbol{\epsilon}_\perp (p,0)=(p_1, p_2), \\
      \epsilon^{\alpha} (p,+1) & = - \left(
      0, \frac{\sqrt{2} (p_1 + i p_2)}{p^+} , \boldsymbol{\epsilon}_\perp (p,+1) \right)^\text{T} \quad \text{with} \quad \boldsymbol{\epsilon}_\perp (p,+1)=(\frac{1}{\sqrt{2}}, \frac{i}{\sqrt{2}}),\\
      \epsilon^{\alpha} (p,-1) & = \left(0, \frac{\sqrt{2} (p_1 - i p_2)}{p^+}, \boldsymbol{\epsilon}_\perp (p,-1) \right)^\text{T} \quad \text{with} \quad \boldsymbol{\epsilon}_\perp (p,-1)= (\frac{1}{\sqrt{2}}, \frac{- i}{\sqrt{2}}).
    \end{align}
\end{subequations}
The massive positive energy Dirac spinor can be written as 
\cite{Lorce:2017isp}
\begin{equation}\label{diraca3}
    u (p, \sigma) = \frac{\left( \slashed{p} + 
    M \right)}{\sqrt{2 p \cdot n}}
 \slashed{n} \chi_{\sigma},
\end{equation}
where $\chi_{\sigma}$ is the rest frame spinor. Note that 
the Rarita-Schwinger spinor in Eq.~\eqref{RSspinor} satisfies the Rarita-Schwinger 
equation, as well as the subsidiary constraint equations,
\begin{equation}
    \left(\slashed{p}-M\right) u^{\alpha} (p, \lambda)=0,\quad
    \gamma_\alpha  u^{\alpha} (p, \lambda)=0, \quad
    \partial_\alpha  u^{\alpha} (p, \lambda)=0.
\end{equation}

\section{On-shell identities}
\label{appendixb}
\vspace{0.2cm} \par\noindent\par\setcounter{equation}{0}
\renewcommand{\theequation}{B\arabic{equation}}

Some useful on-shell identities have been given in Refs.~\cite{Fu:2022rkn}:

\begin{equation}\label{unB1}
  n^{\{ \alpha' \nobracket} {P^{\nobracket \alpha \}}}  i \sigma^{n q}
  \doteq P \cdot n g^{\alpha' \alpha} i \sigma^{n q} + 2 P \cdot n
  P^{\{ \alpha' \nobracket} n^{\nobracket \alpha \}} + \frac{1}{2} (q
  \cdot n)^2 g^{\alpha' \alpha} + 2 q \cdot n P^{[ \alpha' \nobracket}
  n^{\nobracket \alpha ]} + t n^{\alpha'} n^{\alpha},
\end{equation}
\begin{equation}\label{unB2}
    n^{\{ \alpha' \nobracket} P^{\nobracket \alpha \}} \doteq - M \left( 1 - \frac{t}{4 M^2} \right) g^{\alpha' \alpha} \left( \frac{P \cdot n}{M} +
   \frac{i \sigma^{n q}}{2 M} \right) + g^{\alpha' \alpha} P \cdot n +
   \frac{2}{M} P^{\alpha'} P^{\alpha} \left( \frac{P \cdot n}{M} + \frac{i
   \sigma^{n q}}{2 M} \right),
\end{equation}
in which $\doteq$ represents the relation in a similar manner as Gordon identity.
Analogously, for the case of polarized GPDs, one can derive these on-shell identities:
\begin{equation}\label{poB3}
  \begin{aligned}
    4 M n^{[\nobracket \alpha'} P^{\nobracket \alpha]}  \slashed{n} \gamma^5
    \doteq & [4 (P \cdot n)^2-(q \cdot n)^2] g^{\alpha' \alpha} \gamma^5 - 8 (P \cdot n) n^{\{ \alpha' \nobracket} P^{\nobracket \alpha \}} \gamma^5 + 4 (q \cdot n) n^{[\alpha' \nobracket} P^{\nobracket \alpha]} \gamma^5 \\
    & + 8 P^2 n^{\alpha'} n^{\alpha} \gamma^5 + 2 M (q \cdot n) g^{\alpha' \alpha} \slashed{n} \gamma^5,
  \end{aligned}
\end{equation}
\begin{equation}\label{poB4}
  n^{[\nobracket \alpha'} P^{\nobracket \alpha]} \gamma^5 \doteq \frac{1}{2}
  g^{\alpha' \alpha} (q \cdot n) \gamma^5 - M g^{\alpha' \alpha} \slashed{n}
  \gamma^5 + \frac{P^2}{M} g^{\alpha' \alpha} \slashed{n} \gamma^5 - \frac{2}{M} P^{\alpha'} P^{\alpha} \slashed{n} \gamma^5.
\end{equation}

\bibliographystyle{unsrt}
\bibliography{refGPDs}

\end{document}